# A Threat-Intelligence Driven Methodology to Incorporate Uncertainty in Cyber Risk Analysis and Enhance Decision Making

Martijn Dekker
*University of Amsterdam (UvA) - Amsterdam Business School*
*Faculty of Economics and Business*
Amsterdam, The Netherlands
m.dekker4@uva.nl

Lampis Alevizos
*University of Central Lancashire (UCLan)*
*School of Computer Science – Laboratory of Security and Forensic Research (SAFeR)*
Preston, United Kingdom
lampis@redisni.org

*Abstract* — **The challenge of decision-making under uncertainty in information security has become increasingly important, given the unpredictable probabilities and effects of events in the ever-changing cyber threat landscape. Cyber threat intelligence provides decision-makers with the necessary information and context to understand and anticipate potential threats, reducing uncertainty and improving the accuracy of risk analysis. The latter is a principal element of evidence-based decision-making, and it is essential to recognize that addressing uncertainty requires a new, threat-intelligence driven methodology and risk analysis approach. We propose a solution to this challenge by introducing a threat-intelligence based security assessment methodology and a decision-making strategy that considers both known unknowns and unknown unknowns. The proposed methodology aims to enhance the quality of decision-making by utilizing causal graphs, which offer an alternative to conventional methodologies that rely on attack trees, resulting in a reduction of uncertainty. Furthermore, we consider tactics, techniques, and procedures that are possible, probable, and plausible, improving the predictability of adversary behavior. Our proposed solution provides practical guidance for information security leaders to make informed decisions in uncertain situations. This paper offers a new perspective on addressing the challenge of decision-making under uncertainty in information security by introducing a methodology that can help decision-makers navigate the intricacies of the dynamic and continuously evolving landscape of cyber threats.**

## I. INTRODUCTION

The information security domain is rapidly evolving and nowadays already revolves around the notion of uncertainty management [1]. This was highlighted with the update of risk definition within ISO 31000:2018 now stated as "*the effect of uncertainty on objectives*" [2]. This notion was later incorporated to information security domain and ISO 27005:2022 [3]. The same standard provides a framework for managing information security risks, including cyber risks. It also provides guidance on how to identify, assess, and treat risks. The standard has been updated to include the concept of uncertainty as an integral part of the risk management process, recognizing that it is not always possible to predict and prepare for all potential risks.

The inclusion of uncertainty in the ISO 27005:2022 standard highlights the importance of being flexible and adaptable in cyber risk management. It emphasizes the need for organizations to have an initiative-taking approach and to continuously monitor and review their security controls and procedures. Additionally, the inclusion of uncertainty in the standard also implies that organizations should have a clear understanding of the difference between risk and uncertainty and should have different, though intersecting, strategies to manage each.

Risk is a potential event that can be identified and quantified, and its likelihood and impact can be estimated. It can be managed by implementing controls and mitigation strategies that reduce its likelihood or impact. Organizations may use risk management frameworks, such as ISO 27005 [4] or NIST CSF [5] to identify, assess, and treat risks, and to make informed decisions about how to allocate resources to manage them. These frameworks include comprehensive approaches to manage risks, and include key components such as, risk identification, risk assessment, risk treatment and risk monitoring and review. Risk assessment and especially risk calculation is an important part of such frameworks because the results are being used by decision makers to inform the risk treatment process, and thereby make decisions to allocate resources. Consequently, the risk estimation problem is reduced to a well understood financial cost-benefit problem, weighing the cost of the mitigation against the value of impact reduction.

Uncertainty, on the other hand, is the state of not being able to predict or estimate the likelihood or impact of an event. In simple words, the probability distribution of likelihood and impact are unknown. It arises when there is lack of information, or when the information available is incomplete or ambiguous, therefore managing uncertainty is a challenging task. Uncertainty is an inherent aspect of cyber risk management, and it can have a significant impact on the evaluation of cyber security controls during risk assessment. Subsequently, it will impact decision making, both in resource allocation but also in the trust placed in security controls. This is true in many domains, but in the security domain this impact is even more relevant as, due to agency problems, security risks are often over-estimated or under-estimated by non-specialist stakeholders. [6].

Given this distinction however, it is important for organizations to have different strategies to manage each. Previous research on managing cyber risks and decision making has focused primarily on understanding the impact of cyber attacks, the ways to prevent them, and the overall risk management process [7] [8]. It is important for organizations to develop and implement effective cyber risk management strategies that align with modern risk analysis approaches that consider uncertainty [9]. This work examines existing risk assessment analysis orientations and presents a systematic and rigorous threat-intelligence based methodology that builds on existing concepts, however, recognizes uncertainty as an inherent parameter, and thereby aligns with the modern risk definition [3]. Accordingly, decision makers are empowered with information to navigate through uncertainty retaining the ability to course correct, and steer their cyber defences based on the current threat landscape denominated by their IT landscape-specific security controls effectiveness.

The primary motivation for this paper stems from the increasing importance of decision-making under uncertainty in the field of information security. With the dynamic and ever-evolving cyber threat landscape, decision-makers face unpredictable probabilities and effects of events, making risk analysis and mitigation challenging. The inclusion of uncertainty in the ISO standards emphasizes the need for organizations to adapt their risk management strategies. Existing research has focused on understanding cyber-attacks and risk management processes but has paid limited attention to addressing uncertainty explicitly. Therefore, this paper seeks to fill this gap by proposing a cyber threat-intelligence driven methodology that acknowledges and addresses uncertainty, providing practical guidance for information security leaders to navigate the complexities of the evolving cyber threats and make informed decisions in uncertain situations.

The contributions of this paper can be summarized as follows.

**(1) Introduction of a Threat-Intelligence Based Methodology:** we propose a new methodology for cyber security assessment and decision-making under uncertainty in information security. The methodology is driven by strategic, tactical and operational cyber threat intelligence and incorporates causal graphs as an alternative to traditional attack trees. By doing so, it aims to enhance the quality of decision-making and reduce uncertainty.

**(2) Consideration of known unknowns and unknown unknowns:** we can navigate the uncertainty and risk sphere in a way that allows for both traditional and modern risk analysis and approaches to be used more efficiently. As a result, addressing these different types of uncertainties, the methodology provides a comprehensive approach to decision-making.

**(3) Improved Risk Analysis and Predictability:** we consider tactics, techniques, and procedures that are possible, probable, and plausible, improving the predictability of adversary behavior. This enhances risk analysis accuracy and enables information security leaders to make more informed decisions based on the potential threats they may face.

**(4) Practical Guidance for Decision-Makers:** The paper offers practical guidance for information security leaders to navigate uncertain situations. It provides a new perspective and methodology that can help decision-makers understand and anticipate potential threats in the dynamic and continuously evolving cyber threat landscape.

**(5) Alignment with Modern Risk Analysis Approaches:** The proposed methodology aligns with modern risk analysis approaches that consider uncertainty. It recognizes the evolving definitions of risk within ISO standards and emphasizes the need for organizations to have flexible and adaptive risk management strategies.

The rest of this paper is organized as follows: section II. B provides the context, critically discusses existing and related work, and introduces the risk paradox highlighting the problem statement. Section III. Threat-Intelligence Based Security Assessment (TIBSA) details a methodology with unique characteristics that considers uncertainty during risk analysis. Section IV. Decision Making Strategy outlines the key decision-making points within the proposed methodology and highlights the importance for decision makers and modern cyber defences evaluation strategies. Finally, we summarize and draw conclusions in section V. Conclusion.

## II. Background

In this section we discuss the existing relevant approaches to cybersecurity risk assessment. Typically, a risk assessment methodology in the cyber domain includes the following steps: (i) assessment process: the objective is to prepare and establish the assessment context. Identify the purpose, scope, assumptions, constraints, sources and potential risk model and approaches to function as input for next steps (ii) risk model: define key terms and assessable risk factors as well as the relationship between those factors (iii) assessment approach: define the range of values for risk factors and how to analyse the combinations to eventually produce an evaluation of risk with either a numeric expression, severity level or hybrid. Thus, assessment approaches can be quantitative, qualitative, semi-quantitative (iv) analysis approach: define the risk factors and their combinations through several initiation angles. Hence, angles can be asset/impact oriented, vulnerability-oriented, or threat-oriented [9].

There are several globally recognised and used frameworks such as ISO 27005:2022 [4], NIST SP 800-30 [9], COBIT 2019 [10], CIS RAM [11], as well as other frameworks, techniques and methodologies proposed by numerous scholars during their works, nonetheless, each may have slightly varying steps depending on the underpinned risk management framework. We begin, focus, and analyse the existing assessment and analysis approaches with the aim to understand and highlight advantages and gaps.

## A. Assessment Approaches

### 1) Qualitative

A qualitative approach evaluates the effectiveness and efficiency of security measures by using subjective methods. This approach is typically used to assess the quality or level of risk associated with a particular security measure, rather than to quantify the precise level of risk. Techniques usually include interviews with subject matter experts and several stakeholders, review of security policies and procedures, and observation of security practices in action. The goal of a qualitative assessment is to identify strengths and weaknesses in the current security posture and make relevant recommendations. It is worth noting that by gathering subjective information through interviews and observations, a qualitative assessment can provide insights and context that may not be captured by more quantitative methods, therefore such approaches allow for a more in-depth and holistic understanding of the security posture of an organization.

A potential limitation is that the results of a qualitative assessment may be subjective and open to interpretation. This can make it difficult to compare the results of a qualitative assessment to other methods or to other organizations.

Qualitative approaches can be a useful complement to a more quantitative approach, as it allows for a more nuanced and comprehensive understanding of an organization's security posture.

### 2) Quantitative

A quantitative approach evaluates the effectiveness and efficiency of security measures by using numerical or statistical methods. This approach is typically used to quantify the level of risk associated with a particular security measure. The techniques in this case frequently include statistical analysis of security data, such as the frequency and severity of security breaches, or the use of mathematical models to evaluate the likelihood and impact of several types of security risks. The goal of a quantitative approach, contrary to a qualitative one, is to objectively measure the level of risk associated with a particular security measure, and to make recommendations for improvement based on that measurement, which is oftentimes expressed as percentages, value ranges or specific numeric values. By using numerical data and statistical analysis, a quantitative assessment can provide a more precise and accurate understanding of an organization's security posture.

On the other hand, one potential limitation is that it may be difficult to quantify certain aspects of an organization's security posture, such as the effectiveness of its security policies or the level of training and expertise of its security personnel, or assign accurate numeric values in case of e.g., potential loss of data. Additionally, a quantitative assessment may not provide the full context needed to fully understand the root causes of security risks or the potential impact of different risk mitigation strategies.

### 3) Semi-Quantitative

A semi-quantitative or hybrid approach combines both qualitative and quantitative methods to evaluate the effectiveness and efficiency of security measures. This approach aims to provide a more comprehensive and nuanced understanding of an organization's security posture, by combining the in-depth, subjective insights provided by qualitative methods with the objective, numerical data provided by quantitative methods. Techniques include a combination of the two previously discussed methods, and the recommendations for improvements are based on both qualitative and quantitative data. Semi-quantitative approach allows for a more complete and well-rounded understanding of an organization's security posture. By combining both qualitative and quantitative data, a semi-quantitative assessment can provide a more comprehensive view of the security posture and can help to identify areas for improvement that might not be identified using a single method.

A potential limitation, however, is that it may be more time-consuming and resource-intensive than a purely qualitative or quantitative approach. Furthermore, combining different data sources and methods can be challenging, and may require specialized expertise to interpret the results accurately and effectively.

## B. Assessment Analysis

### 1) Vulnerability-Oriented

This orientation starts with the identification of exploitable weaknesses in organizational assets, or the ecosystem in which they operate. Threat events that could exploit those vulnerabilities coupled with potential consequences of vulnerabilities being exploited are then analysed. This orientation is grounded upon glossaries that classify vulnerabilities such as Common Vulnerabilities and Exposures (CVE) [12]. Common Vulnerability Scoring System (CVSS) [13] is used to evaluate the threat level of each vulnerability. Conclusively, a vulnerability catalogue is created and maintained e.g., the NIST's National Vulnerability Database (NVD) [14] and used as the foundation.

Northern et al. [15] presented a CVE-based methodology for cyber physical systems. Their methodology, when combined with the controlled moving target defence concept which immediately replaces the identified vulnerable component, can improve resilience against cyber-attacks. However, it is inherently subject to a set of predisposing conditions and uncertainties which are not considered e.g., unrecognized vulnerabilities, unrecognized dependencies, and thereby unforeseen impact. George and Thampi [16] proposed a vulnerability-based risk assessment towards edge devices on the Internet of Things (IoT) based on CVSS. Their work starts with vulnerability identification through a vulnerability scanner, such as Retina IoT Scanner, IoT Sploit, or Kaspersky IoT. Next, they used attack graphs to calculate the probability of attack based on the vulnerability values found in attack paths from edge devices towards the target IoT system, semi-qualitatively evaluating the risk, and proposing mitigating measures. Aksu et el. proposed a quantitative CVSS-based methodology while leveraging attack graphs [17], however, they considered low level metrics such as complexity, capability, exploitability, contrary to the high-level focused works previously seen. Ushakov et al. [18] developed a risk assessment technique and tool based on CVE and Common Platform Enumeration (CPE). Their algorithm supports the mapping of vulnerabilities to software and automates the searching against known vulnerabilities through NVD, nevertheless, it does not consider threat-intelligence sources. Russo et al. [19] presented an automated vulnerability-oriented approach through a custom-made software platform, based on NIST 800-30 guidance [9].

Contrary to our work, these techniques and methods are subject to the same predispositions addressed earlier in this section. Namely, there is little to none provisioning for uncertainty management, as they do not consider (1) threat-intelligence or threat information in general, and (2) threat source's behaviour, thereby potentially leading into narrow conclusions, which would in turn leads to narrow security measure insights. Thus, ultimately resulting in single faceted decision making rather than multi-faceted, hence the outcome is a single faceted cyber defence strategy unable to cope with emerging threats [20].

*2) Asset/Impact-Oriented*

The asset/impact-oriented approach begins with: (1) the identification of impacts on critical assets e.g., via business impact analysis, and (2) the identification of threat events that could elicit those impacts.

Li et al. [21] presented a dynamic asset/impact assessment approach for modern industrial control systems (ICSs). Assets receive specific attribution and then a trend of impact is dynamically predicted based on the aggregated information on a subject asset. The overall impact is quantitatively measured based on specific values and properties per asset. The authors provide a comparative study with other similar asset/impact approaches, highlighting some differentiating characteristics of their own work. For instance, the quantification ability, the prediction of trend of impact and the impact propagation analysis. Although this work emphasizes on the importance and criticality of both assets and impact respectively, it lacks insights such as threat source's attribution and objectives, which affect the overall risk evaluation. While the uncertainty factor is being considered, it is seen as explicitly single-sided from the impact perspective, hence this is contributing towards a reactive damage control rather than a balanced proactive-reactive cyber defence strategy. Kure & Islam [22] introduced a semi-qualitative asset/impact focused approach based on NIST SP800-30 [9] and ISO 31000 [23] to assess risks against Critical Infrastructure (CI). The drawback of this approach is the generic threat catalogue used to identify the impact per asset, and the checklist mindset when identifying vulnerabilities. Rea-Guaman et al. [24] proposed a new asset/impact-based approach, supported by their own risk management framework named AVARCIBER. This work quantitatively measures impacts and provides feedback to the overall risk management framework to support decision making. Resembling the previous work however, it is based on generic threat and vulnerability taxonomies provided by ISO 27005 [25] and ENISA [26] respectively. Generic taxonomies lead to generic control assessment, thereby providing equally generic conclusions to decision makers.

*3) Threat-Oriented*

The threat-oriented approach begins with the identification of threat sources and their respective threat events. Contrary to the asset/impact-oriented approach, the vulnerabilities are identified in the context of threats, while the impacts are identified based on the adversary's intention.

Kim & Cha [27] devised the Security Risk Analysis (SRA), a qualitative method to address risks starting with the development of threat scenarios. Although threat scenarios form a solid foundation to begin with, their next step is to continue the scenario breakdown and threat event build up with an internal group of stakeholders. As a result, not only the actual threat landscape is at risk to be missed, but the threat event list is drawn in a heavily subjective manner. Haastrecht et al. [28] introduced a threat-based assessment analysis combined with metrics acquired from users and devices. The authors draw a conceptual data model that correlates all relevant data from endpoints (user devices), acting as input for their algorithm to calculate risk scores per user and device according to threats. The proposed source of threats and threat events however, is the annual ENISA report of top cybersecurity threats, which remains remarkably unchanged through the years, with ransomware being the exception/addition [26] [29]. That said, firstly, the current threat landscape is evolving as fast (if not faster) as the global information technology (IT) landscape, thus, the annual cadence of threat sources and events monitoring is likely insufficient. Secondly, this approach is positioned for user-endpoints, hence leaving out of the equation important IT building blocks and thereby providing limited insights for holistic decision making. Haji et al. [30] proposed an improved method based on attack trees and the open web application security project (OWASP) [31] for the scenario construction. Ahmed et al. [32] extended and improved the previous works even further. They introduced a quantitative MITRE ATT&CK driven approach based on NIST 800-30 [9] principles and attack trees. Nonetheless, both [30] and [32] are subject to similar predispositions. Specifically, the former utilized OWASP, which provides industry leading insights on threat events related to web applications, services, or application programming interfaces (APIs), while the latter used ATT&CK [33] which is considered as the industry leading knowledge base for threat events. Furthermore, using either of the two mentioned threat knowledge basis in explicit mode, combined with attack trees comes with an inherent drawback. Namely, attack trees being purely hierarchical structure do not realistically reflect the adversary's perspective, compared to e.g., attack graphs or causal graphs that allow for more relaxed root-node-leaf connections, hence a more realistic adversary behaviour. Consequently, they could provide potentially good insights on risks, however, they do not consider uncertainties from the adversary's perspective and their motivations.

## *C. The Risk Paradox*

Oftentimes risk analysis may provide unexpected insights about how the actions of defenders can impact the attackers. It is natural to think about the attack process from the perspective of defender, but it is important to remember that the two parties involved in the situation have different and usually vastly unrelated goals. It is essential to consider the interests of the defender, but it is imperative to recognize that the interests of the attackers and the defenders are completely detached. Thus, the attacker's decisions are based solely on their own interests and do not consider the effects on the defenders. This is a fundamental differentiator when conducting risk analysis, and more importantly, when providing insights to decision makers either regarding the resilience of the security defences, or the operational effectiveness of the security measures.

To practically understand the magnitude of this differentiating factor, consider the following example: threat-intelligence (TI) sources provide insights of an adversary performing ransomware attacks via two specific scenarios, described via tactics, techniques, and procedures (TTPs) that have been observed and verified by TI to have high propensity and success ratio. As such, the distribution of the TTPs in the subject scenario are known, hence they pose a risk for the defender. Both scenarios have high propensity, but one of the two has slightly higher success ratio than the other, hence chosen by the adversary. The two scenarios correspond to different leaves in an attack tree [34], or in general, they represent different attack paths in a threat model. A security analyst identifies this and implements a security control to lower the likelihood and impact of TTPs. The adversary responds with the next best scenario but with slightly lower propensity, however, due to specific circumstances within the defender's IT landscape (e.g., technological incompatibilities) this is not applicable. The adversary, however, can overcome this specific obstacle by e.g., porting libraries that will deem the scenario applicable again, selecting a different TTP that was not considered probable, or even, improvising a novel TTP. Strangely, although the likelihood and impact of the attack has significantly decreased after the implementation of the security control, the overall risk has significantly increased. Similarly, actions by defenders to block a specific TTP, can motivate the attacker to switch to another TTP that turns out for example to be less detectable by the defender, resulting in an increase of impact. This demonstrates the relevance to include the interplay between attack and defence actions in your risk analysis, but also shows that this introduces uncertainty.

Convincingly, as we accentuate with our methodology in this paper, it is imperative to understand that: (1) adversaries' TTPs may have a known probability distribution, however, they are heavily influenced by the defender's actions and the specific IT landscape where they unfold. (2) adversaries' actions and reactions include considerable uncertainty; thus, some TTPs probability distribution is unknown. (3) defenders must consider plausible and probable scenarios and TTPs, on top of possible ones, to augment risk analysis with uncertainties. Security assessment is a critical element of evidence-based decision making; hence it is essential to enable decision makers to manage risks, and uncertainties. Risk treatment should therefore include ways of reducing the uncertainty in the probability distributions. Modern decision makers in the cyber security domain with the goal of orchestrating cyber defences with increased resilience, need to manage not only the known-unknowns but also unknown-unknowns.

## III. THREAT-INTELLIGENCE BASED SECURITY ASSESSMENT (TIBSA)

### *A. Methodology*

This section details the TIBSA, which is a methodology with two core objectives: foster interoperability amongst various IT, security and other capabilities, and support decision makers in building resilient cyber defences both under certainty and uncertainty. TIBSA may be performed at its fully fledged form, but there can also be a rapid-TIBSA version. Meaning that the level of rigor applied in TIBSAs can be scaled up or down accordingly. For instance, when dealing only with known unknowns (e.g., TTPs probability distribution can be defined) those can be moved into the sphere of risk and thereby several standard analyses methods can be applied, including the rapid-TIBSA. Nevertheless, when dealing with unknown unknowns and thus uncertainty, (e.g., TTPs probability distribution is not known) then TIBSAs core aspects (ref. Figure 1) will help achieve significantly better results, and therefore superior decision making.

#### *1) Foster Interoperability*

TIBSA fosters interoperability by intersecting the various capabilities within an organisation's security functions, e.g., capabilities belonging to identify, protect, detect, respond, and recover layers [5]. The times where decision makers invested and implemented numerous security products to tackle various security issues, which led into a technological burden and introduced incompatibilities while wasting effort and resources, are not sufficient anymore. The same analogy applies to people and processes. Interoperability must be achieved both on technology and process level for the decision makers to build modern, resilient defences.

#### *2) Support Decision Making*

TIBSA enables decision makers to identify, prioritize, and respond to cyber threats, through the evaluation of effectiveness of security controls and their implementation, eventually reducing susceptibility against cyber threats. Several capabilities aim to prevent or detect for example, can be improved by either technical or administrative fine tuning. Effective security defences do not always demand more security controls, and more security controls do not automatically imply effective defences. It is for the organization's ability to provide the right amount and quality of information to decision makers, which makes of an effective defence ultimately.

The core aspects of TIBSA methodology are shown in Figure 1, and discussed in detail in the next section.

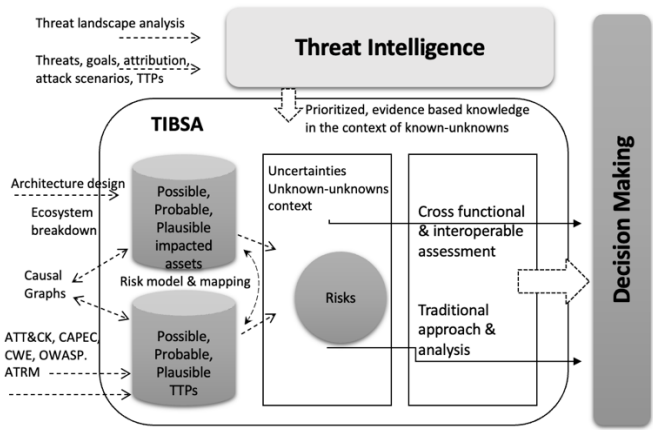

*Figure 1 - Core aspects of TIBSA on high-level.*

## B. Approach and Analysis

TIBSA evaluates an asset's ability or inability to withstand cyber-attacks over a range of TTPs, in a semi-quantitative manner. Although the primary source of information is the threat intelligence capability, other source of inputs can be considered subject to each organisation's maturity. For instance, in some organizations where threat-based or risk-based IT audit approaches are adopted, they can provide potentially insightful data to tailor and facilitate an ad-hoc TIBSA from a rather technical perspective. TIBSA consists of the following steps: (1) understand the cyber threat landscape (2) identify possible, probable, and plausible impacted assets (3) identify possible, probable, and plausible TTPs (4) apply scoring model (5) identify security controls in place (6) assess control effectiveness.

### 1) Understand the Cyber Threat Landscape

TIBSA necessitates high quality, evidenced-based knowledge such as, threat context, indicators, implications, mechanisms, behaviours, and action-oriented advice by threat-intelligence, as the first step. In other words, understanding the cyber threat landscape begins with a well-established and mature cyber threat intelligence operating throughout strategic, operational, and tactical levels This allows for data collection, processing, and analysis to understand threat source's goals, motives, targets, patterns, behaviours, and attribution [35]. Cyber threat intelligence (CTI) is an enabler for more informed, data-backed security decisions and thereby the first and crucial step to begin the TIBSA.

CTI finds several use cases nowadays. For example, it empowers C-level executives with insights that may help to make faster and more efficient decisions. It also sheds light to potential business-specific threats, and thereby enables security teams to make better decisions, e.g., improving the security posture by prioritizing vulnerability remediation, or calibrate prevention and detection mechanisms. Moreover, strategic, and tactical level CTI capabilities empower other security capabilities by revealing adversarial goals, motives, attributes, modus operandi and specific TTPs [36] and perform rigorous threat research. It is not the goal of this paragraph to thoroughly examine CTI's contribution and specifics within the cyber domain, however, it is rather crucial to establish CTI as the guide for TIBSA, by continuously monitoring and analysing the cyber threat landscape throughout all three strategic, operational, and tactical pillars. Thus, it is imperative for CTI to provide actionable, evidence-based knowledge on potential threats, their goals, and/or their TTPs for TIBSA to initiate.

### 2) Identify Possible-Probable-Plausible Impacted Assets

Starting the assessment from a threat-intel perspective, may seem that the scope is arbitrary, compared to an asset-oriented analysis for example which starts with a standard asset list, however, it is not. Given the CTI feedback, the main activity of this step is to define the scope of the assessment and draft a candidate asset list subject to impact.

This may be achieved by (1) follow CTI's threat research to identify impacted assets led by the threat's goal, attack scenarios and paths, TTPs, and (2) draw tailored scenarios grounded on organization's specific IT landscape and technology stack, with the goal to further refine the list of impacted assets. Note the "and" condition in the above two actions in this step, as it plays a pivotal role in continuation. Specifically, the applicability of TTPs and the testing of effectiveness of relevant security controls will either be increased or decreased based on the results of these actions. Decisively, the scoping or de-scoping of TTPs and their respective security controls will have a significant impact on the decision making for the overall security program of an organization. This in turn translates to either financial or operational security impact.

Collaborative effort in the context of such actions is key and contributes towards a successful and complete list of impacted assets. Consequently, the close collaboration between CTI and TIBSA is of utmost importance, nonetheless, it is equally beneficial to consider the feedback of other relevant capabilities such as red team, threat hunting, and IT architects. Ultimately, the list of impacted assets is a derivative of (1) direct analysis of the provided evidence-based knowledge by CTI (2) goal-based analysis of the threats through causal graphs. Lastly, when drafting the list of impacted assets TIBSA embraces the "assume breach" mindset as the default mode. Therefore, when forming testable hypotheses with the goal to find impacted assets, TIBSA does not rely on the typical threat source differentiation. So, TIBSA does not inherently treats assets residing in the internal part of the network as more secure than external facing assets. This, however, depends on the depth of the assessment, complexity, capacity, resources, expectations, time constraints and other organization specific parameters. Since this is a conscious decision that needs to be taken upfront, when engaging in a rapid-TIBSA there is always the option to fall back to the typical, perimeter-centric approach which differentiates between internal and external zones.

### 3) Identify Possible-Probable-Plausible TTPs

Once agreement or at least consensus is reached amongst stakeholders on impacted assets list with precision and confidence, the next step is to identify possible, probable, and plausible TTPs. This step includes three activities: (1) list of impacted assets may be provided directly by CTI. (2) the use of technical-oriented knowledge base of adversary TTPs, rather than generic threat event catalogues, to identify probable TTPs (3) identification of possible, probable, and plausible TTPs through causal graphs and tailored threat analysis based on organization's specific IT landscape and technology stack.

The first activity is rather simple, a tactical and strategic level CTI provides concrete and trustworthy research that can be consumed immediately. Thereby, attach TTPs directly against the list of impacted assets is the immediate first action.

The second activity, however, involves the identification of probable TTPs through the most suitable knowledge base. This action requires extensive understanding of both the subject asset, as well as the available technical TTP knowledge bases. TIBSA leverages each technical-oriented knowledge base according to the technology specifics and draws the best match correspondingly. For example, MITRE ATT&CK [33], which is considered the "lingua franca" and used by 89% of the organizations according to ESG research [37] is more focused on infrastructure components. Although it includes web application specific TTPs, it is not as in-depth as the infrastructure related TTPs for the time being. It also contains platform specific TTPs, such as Windows, Linux, MacOS, Azure, and others. Another notable example, well-versed for web applications and development lifecycle, however, is MITRE's Common Attack Pattern Enumeration and Classification (CAPEC) [38]. Additionally, there are vendors providing platform-explicit knowledge bases of adversary TTPs, such as Microsoft's Azure Threat Research Matrix (ATRM) [39]. That said, TIBSA is not limited to one specific technical knowledge base of adversary tactics. It utilizes a hybrid approach connecting the TTPs from the highest level of tactics down to specific assets on the previously identified impacted asset list, based on a best-fit approach. Figure 2 presents an example of this activity in the context of TIBSA using a combination of MITRE ATT&CK [33] for tactic, techniques and sub-techniques, MITRE CAPEC [38] for attack patterns, MITRE Common Weakness Enumeration (CWE) [40] for the weaknesses, CVE [12] for the vulnerabilities, and vendor specific resources for the assets.

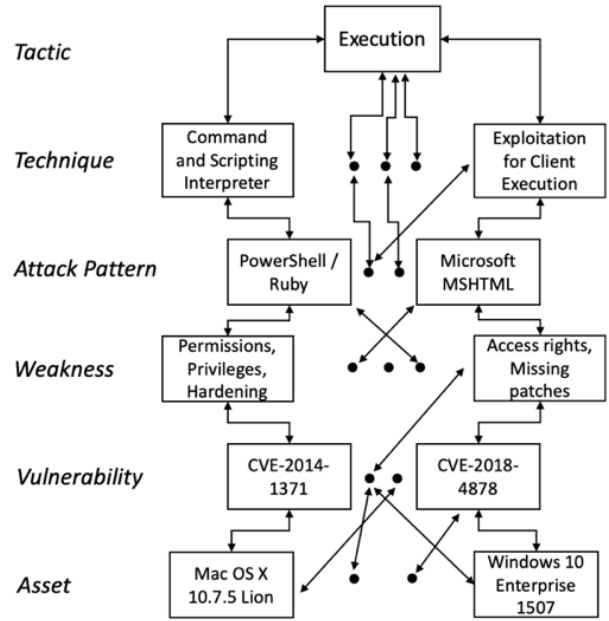

*Figure 2 - Paths related to execution tactic.*

The third activity involves the identification of probable TTPs using causal graphs. At this step, this activity is usually combined to complete the causal graph drawn during the previous step, thereby having the full cause and effect relationship between impacted assets and TTPs. Drawing causal graphs used to be a complex undertaking, however, the Center of Threat Informed Defence (CTID) [41] recently launched a data model that simplifies this process. Figure 3 shows a causal graph example related to FIN13 campaign targeting banks in Latin America [42].

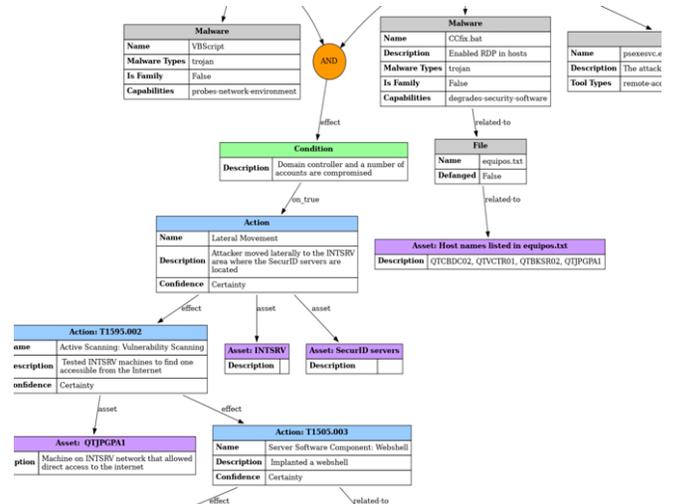

*Figure 3 - Causal graph for FIN13 campaign [41].*

### 4) Apply Scoring Model

TIBSA's design allows and demands for meticulous actions. Nevertheless, not all organizations have the same resources, objectives, mission, and vision. Therefore, where some organizations would opt-in for a full-scale TIBSA, assessing all applicable security controls against the possible, probable, and plausible TTPs, others may find a scaled version (rapid-TIBSA) more suitable. Regardless the chosen TIBSA version, a scoring model is a fundamental step towards prioritization of TTPs coverage.

Scoring models can be applied through several methods. For instance, a scoring model yields remarkable results if applied through the simplest manner, spreadsheets. However, it can also be implemented in an AI-enabled system to facilitate ease of use, automation, potentially reduce the necessity of highly knowledgeable experts, limit subjectivity, or even reduce bias. It is advisable to customize and implement the model in an automated fashion and provide a web user interface. As such, best effort results and the best user experience may be achieved. In principle the model must (1) be customized to reflect organization's needs (2) be tailored to reflect adversary's specific threat models e.g., add weighted factors according to causal graph (3) be validated by multiple assessors to normalize bias where necessary e.g., by considering ranges/average of aggregated score (4) be used consistently to allow for valid and relative comparisons (5) be updated continually. Table illustrates an example model, showing the range of factors versus the criteria range and their linear numeric scoring [1...5].

*Table 1 - Scoring model example.*

| Range of Factors / Criteria Range | 1 | 2 | 3 | 4 | 5 |
|---|---|---|---|---|---|
| Is there **evidence of this TTP** in a reputable adversary knowledge base? | No evidence of TTP | Scattered information / possible use of TTP | Confirmed evidence of TTP in at least one knowledge base | Confirmed evidence of TTP plus frequent use reported | Confirmed evidence of TTP plus widespread use reported |
| What is the **level of skill** required to apply this TTP? | Advanced skills and specific knowledge on the targeted system | Advanced skills on the targeted asset | Some skills on the targeted asset | General technical skills | No specific skills required |
| What is this TTP's **applicability**? | Single asset | Small number of assets system in isolated zone with monitored internet access | Entire ecosystem | A system of systems | A significant portion of IT landscape |
| What is the **positioning effect** of this TTP? | General non-segmented, non-monitored network with internet acces | General non-segmented network with internet access | General segment with internet acccess | Isolated zone with internet access | Isolated zone with no internet access |
| How long would it take to **recover** from this TTP once detected? | <8 hours | 8-16 hours | 17-37 hours | 38-52 hours | > 52 hours |
| What is the **estimated cost** to restore or replace the impacted asset? | < 10k € | 25k € | 50k € | 75k € | > 100k € |
| How detectable is this TTP when applied? | TTP obvious without monitoring | Detection likely with routine monitoring | Detection likely with simple refinements of detection methods | Detection possible with newly introduced detection methods | Undetectable |
| What is this TTPs **confidence level** assigned in **causal graph**? | Extreme uncertainty | Large uncertainty | Certainty | Large certainty | Extreme Certainty |

*5) Identify Security Controls in Place*

Security controls are safeguards or measures prescribed for an information system or an organization, designed and implemented to protect the confidentiality, integrity, and availability of its information and to meet a set of defined security requirements [43]. One of the goals of TIBSA is to assess operational effectiveness of the applied security measures, or otherwise referred to as security controls. Operational in this context means controls that are in place and active in production environment that should be explicitly considered. Next, controls must be mapped to possible, probable, and plausible TTPs.

To characterize the effectiveness of each control we define four primary criteria (1) prevent: controls aim to prevent TTPs execution (2) detect: controls aim to uncover the presence or actions of a TTP (3) constrain: controls aim to reduce the risk associated with TTPs e.g., by reducing the likelihood or severity parameters (4) recover: controls aim to recover from either a TTP or an attack. Above criteria consist of the foundation range only. Organizations are advised to customize this range accordingly. For example, organizations with mature cybersecurity programs may utilize pre-emptive security controls, whereas others employ deception controls. Subsequently, organizations should customize criteria according to the nature of their in-use controls, while considering the context and the specific technologies operating in their specific IT landscape.

*6) Assess Security Controls in Place*

TIBSA is designed with interoperability in mind. It fosters collaboration between security capabilities regardless of their positioning within an organization. For example, cross-division capabilities may form virtual teams tasked with the same goal, hence allowing for breaking potential silos. This in turn not only boosts collaboration, but allows for various expert insights consolidation, thus forming significantly refined, and as much as possible, bias free conclusions. It is therefore important to assign the best fit-for-purpose capability to evaluate a security control's effectiveness, corresponding to TTPs. Assessment of controls via technical workshops and interviews may be assigned to assessors, while controls demanding in-depth technical validation may be assigned to technically savvy experts such as penetration testers. TIBSA may have an impactful cooperation with threat-intelligence based ethical red teaming (TIBER) during control assessment. As described in the TIBER-EU framework [44], TIBER performs a TI driven capture the flag exercise. Hence, TIBSA may involve TIBER where applicable as a precision test for a range of TTPs. On the other hand, TIBER may provoke broader, ecosystem based TIBSAs. Assertively, to achieve effective collaboration and clearly scoped work distribution, a TTP versus in-place controls mapping is necessary, and thus, the first activity of this step. Effectiveness may have several levels of granularity, however, hence up to organizations to define their own according to needs and maturity levels.

*Table 2 - Mitigating criteria and effectiveness scales.*

| Effectiveness | Mitigating Criteria & Scoring | | | |
|---|---|---|---|---|
| | **PREVENT** | **DETECT** | **CONSTRAIN** | **RECOVER** |
| **High** | PR.H = 12 | DT.H = 8 | CS.H = 7 | RE.H = 5 |
| **Medium** | PR.M = 10 | DT.M = 6 | CS.M = 5 | RE.M = 3 |
| **Low** | PR.L = 8 | DT.L = 4 | CS.L = 3 | RE.L = 1 |

Table 2 shows an example of effectiveness scale versus pre-defined criteria. TIBSA uses the following two letter notations, inspired by [45] and grounded on the example criteria (prevent, detect, constrain, recover) to help simplify and speed up the completion of this activity. The third letter (L, M, H) signifies the degree of effectiveness. For example, some controls may be highly effective in preventing a TTP but provide for low or even zero recovery value. Others may be highly effective in detecting TTPs and moderately effective in constraining a TTP. The objective of this step, however, is to assess in-depth and conclude on the in-use security controls effectiveness against the range (possible-probable-plausible) of TTPs.

TIBSA implements a simple yet effective and pragmatic approach to conclude on the effectiveness evaluation of an in-use control, based on the principles of benefit-cost analysis (BCA) [46]. A linear scale [1…12] is assigned as shown in Table 2. It is worth noting that the score decreases from left to right, where left means prevention controls are valuated inherently higher than recovery controls. As a result, prevention-first strategies would be favoured overall compared to reactive and recovery-first strategies, nonetheless this is again subject to organizational needs and can be adjusted accordingly. To calculate the first factor, benefit, equals to the summing of scores against the range of mitigated TTPs based on Table 2. To calculate the second factor, cost, parameters such as cost to develop, implement, or maintain a security control must be considered and aggregated. In Table 3 we bring it all together to show the in-use controls mitigating effectiveness against the range of TTPs, arranged with decreasing B/C ratio. As discussed earlier (see 4) Apply Scoring **Model**), different approaches and potentially more complex ones with weighted criteria can be implemented subject to organization's maturity and needs.

*Table 3 - TTPs vs in-use controls and B/C ratio.*

| | Control ID | TTPs IDs | | | | | | | Benefit | Cost | B/C Ratio |
|---|---|---|---|---|---|---|---|---|---|---|---|
| | | T1134 | T1087 | T1110 | T1059.001 | T1059.007 | T1078 | T1562.001 | 12 | 1 | 12 |
| In place security controls - Mitigation Effectiveness Matrix | ST7.C098 | PR.H DT.H | PR.H DT.H | | CS.M | CS.L | | | 11 | 1 | 11 |
| | ST6.C121 | PR.H | | DT.M | DT.M RE.L | | | | 11 | 1 | 11 |
| | ST1.C007 | | | CS.M | | | RE.L | RE.L | 18 | 2 | 9 |
| | ST5.C051 | | DT.M CS.M | PR.L DT.L | RE.H | | | PR.M CS.M | 16 | 2 | 8 |
| | ST9.C101 | | | | | | RE.H | | 16 | 3 | 5.3 |
| | ST5.C054 | | DT.H | | | CS.H DT.H | | | 10 | 4 | 2.5 |
| | ST3.C038 | PR.L DT.H | | PR.L DT.H | CS.M | | RE.M | RE.H | 7 | 3 | 2.3 |

#### 7) Reporting and Recommendations

The last step of TIBSA, is to document the results e.g., in a central risk register, to enable proper follow up and risk/issue management activities. Recommendations are critical and must be translated into well-formed, factually supported technical observations, as well as executive level language.

This step includes the following activities, the output of which must be summarized within the report and recommendations: (1) since controls may be assessed by different teams, experts, or capabilities (e.g., TIBER, threat hunting) it is equally important to normalize and consolidate these inputs prior reporting. Nevertheless, there must be consistency throughout the use of metrics and scales to achieve trustworthiness in comparisons. (2) Clear recommendations on effectiveness of controls against initially scoped threat's goal, attack scenarios, or TTPs. (3) Concise recommendations to stakeholders, e.g., assuming the relevant part of report is received by security operations centre (SOC), which refines a detection rule, thus improving effectiveness from DT.L to DT.H and resulting into better B/C ratio. (4) The reasons why controls are effective – ineffective against range of TTPs must be documented, communicated, and ideally factually checked by at least two assessors. (5) The impact of leveraging a potential opportunity (implementing a new control) as well as the impact of adjusting controls to reflect different strategies (e.g., preventive to reactive), must be documented and reported.

### IV. Decision Making Strategy

#### A. The Ellsberg Paradox

The Ellsberg paradox [47] is a thought experiment in decision theory that illustrates how people's preferences for risky options can be affected by the level of uncertainty involved. It is based on a scenario in which a person is presented with two urns, each containing a certain number of red and black balls. In the first urn, the person knows that there are 50 red balls and 50 black balls. In the second urn, the person only knows that there are 100 balls, but the distribution of red and black balls is unknown. The person is asked to choose one ball from one urn, and drawing a red ball means you will get a reward.

The paradox is that, in most cases, people will prefer to choose a ball from the first urn, where the proportion of red and black balls is known, even though the probability of drawing a red ball from the second urn could be higher than 50%.

The Ellsberg paradox highlights the fact that people's preferences for risky options can be affected by the level of uncertainty involved and that people tend to prefer decisions with measurable risks over decisions with unknown risks, even when the reward can be lower following such strategy. This can have important implications for decision-making, particularly in situations where there is a high degree of uncertainty, and the probabilities are unknown.

To deal with the uncertainty of the Ellsberg paradox, one could ask to take a sample of both urns first and use that information to reduce the uncertainty of the probability distributions. using the key game theoretic concept of rational solution [48]. This is a novel concept particularly designed for coherent or quasi-coherent games, in which every player engages in counterfactual reasoning. Which means that each player knows or, at least, believes that his own action may affect the actions chosen by his opponents. Ultimately, every player has an idea of how his own actions influence the actions of the others. Contrary to the Nash equilibrium [49] the notion of counterfactual reasoning as described in the rational solution may add a valuable perspective in cyber risk analysis.

Similarly, we pose that the TIBSA methodology allows for not only considering uncertainties driven by strategic level CTI and supported by causal graphs, but also for assessing the effectivity of uncertainty reduction of security measures considering counterfactual reasoning while staying within the realm of utility and plausibility.

### B. From Uncertainty to Risk and Vice Versa

TIBSA incorporates uncertainty by default and considers that both the adversaries and defenders will make decisions under a high degree of uncertainty. The Ellsberg paradox briefly explained above, is the perfect analogy to better understand the concepts discussed in step 3) Identify Possible-Probable-Plausible TTPs.

Firstly, we must explain the terms known unknown and unknown unknown in this context. A known unknown refers to a situation where an event is known to happen, but the probability distribution and the specifics of this event are unknown. On the contrary a situation where neither the event nor the probability distribution is known, refers to unknown unknown.

TIBSA begins by receiving evidence-based knowledge through CTI. Strategic level CTI thereby plays a crucial first role conducting analysis on the unknowns of the threat landscape. Provided input may be a threat with TTPs attached to it, or it may be a specific advanced persistent threat (APT) campaign with discrete TTPs in each attack phase. Conversely, evidence-based, and trustworthy CTI can attach probability distribution to threats through rigor analysis, therefore transfer these threat events from unknown unknowns to known unknowns. In other words, transfer those threat events into the risk sphere (see Figure 1). That is, because we know that a threat actor is targeting a specific business industry, the modus operandi is known and the TTPs, thereupon the probability distribution becomes known (remember Ellsberg paradox).

However, defenders do not know how exactly adversaries will react within their specific IT landscape when faced with constrain controls or other unpredictable for them factors, e.g., technology incompatibilities or limitations. Thus, the inherent uncertainty that the modern assessment analysis must consider. TIBSA is not a one-size-fit-all methodology. The uncertainties are considered by design, nonetheless its flexibility allows for opportunities to conduct rapid-TIBSAs through the application of typical assessment analysis. Subsequently the decision making is further enhanced by allowing the choice of operating mode accordingly, which in turn yields the difference between effective and efficient defenders against uncertainty and adversaries.

### C. From Possible to Probable to Plausible

Possible TTPs are those capable of happening, existing, or being true without contradictory facts, laws, or circumstances. Probable TTPs are those likely to happen or to be true, likely but uncertain [50]. Plausible TTPs are those seemingly or apparently valid, given the bounds of uncertainty [51].

An oversimplified example would be the following: consider a financially motivated threat group e.g., FIN7 which leverages REvil[1] ransomware and their own Ransomware as a Service (RaaS) to extort money from the victims [52]. CTI provides analysis stating that one of the TTPs used to compromise a user device, also known as endpoint, is via phishing campaigns containing a malicious document as an email attachment. Up to this point, only known unknowns are being considered, namely, threat events with known probability distribution. In continuation, adversaries used a customized ransomware version specifically designed to encrypt virtual disk volumes of ESXi[2] servers. If this APT campaign is assessed via traditional analysis methods, while the organization is not using ESXi technology but Citrix Hypervisor, the result will place high confidence and trust on the asset's ability to withstand FIN7 [53]. This is not entirely true, however. As we highlighted in this work, existing analysis methods do not account for uncertainty. In this case, the uncertainty is FIN7 reacting to a limitation and devising a new TTP. Viz. reacting to IT landscape organization-specific parameters that are being presented, to achieve their goal. Thus during steps 2) Identify Possible-Probable-Plausible Impacted Assets and 3) Identify Possible-Probable-Plausible TTPs, organizations must consider such factors but always on the basis of what is possible first, and then probability and plausibility within the specific IT landscape. The objective of both steps is to account for uncertainty within the given technological and organizational boundaries, with the goal of threat actor as the denominator, thus avoiding analysis paralysis situations [54].

### D. From Attack Trees to Causal Graphs

Attack trees are branching, purely hierarchical structures that represent ways to achieve an event in which system security is compromised in a specific way [55]. Meaning that nodes in an attack tree may have multiple children but only one parent, with the exception being the root node. Causal graphs on the contrary, are graphical models used to represent the probabilistic relationships between threats, TTPs, vulnerabilities, and assets, and the potential impact of different attack scenarios. A causal graph typically consists of nodes, which represent variables, and directed edges, which represent the probabilistic relationships between variables. For example, a node may represent a specific vulnerability and an edge may represent the relationship between that vulnerability and a specific threat (see Figure 3). It is thereby causal graphs that allow for such uncertainties to be accounted for, contrary to attack trees.

---

[1] https://attack.mitre.org/software/S0496/

[2] https://www.vmware.com/nl/products/esxi-and-esx.html

It is imperative to understand that adversaries may use exploits, malware, and a plethora of tools to pursue their goals, but to do so, they will also use legitimate tools, pretend to act legitimately, and finally, adapt to the target IT landscape. That said, adversaries may be security or IT, however with vastly different goals. To tackle this problem defenders must stop trying to think like attackers explicitly. They must start thinking as attackers but detached from their defender's perspective during assessment analysis [57]. Meaning that defenders adopting the adversary's perspective oftentimes think in lists or trees. However, attackers think their goals. For instance, attackers may land into a trusted zone within an organization's network through spear-phishing, but that does not mean they landed in a node of attack tree. Rather they landed in a node of a causal graph. That is, attackers will move towards their goal by course correcting accordingly, regardless of prescribed actions in a list or a tree. Thus, causal graph is immensely different from the one used mentally and practically by IT and security to manage or monitor the network respectively. The use of causal graphs allows for TIBSA's analysis method to focus primarily on the goal, cause, and effect of adversaries and therefore empower decision making under uncertainty.

## V. Conclusions & Future research

This paper details a flexible and practical threat-intelligence driven analysis method, which considers uncertainty and improves decision making. Incorporating uncertainty into assessment analysis, and specifically in the process of evaluating cyber security control effectiveness, can help chief information security officers (CISOs) to make more informed decisions about resource allocation and how to protect against cyber risks. Considering the level of uncertainty associated with different risks and controls, CISOs can better understand the potential impact of different risks and the effectiveness of existing controls in mitigating those risks. This can help ensure that resources are allocated efficiently and effectively based on organization's needs and that the organization's security posture remains continuously suitable for the emerging threat landscape. Finally, decision makers may be able to avoid overspending by utilizing a cost-benefit approach as proposed, to determine the most cost-effective controls for mitigating the identified risks. This in turn provides trustworthy information and actionable insights to CISOs to avoid the common pitfall of over or under trusting security controls, and thereby fine tune their security defences.

While the methodology provides a valuable framework for decision-making under uncertainty, several research challenges remain. One area of future research is to focus on refining and enhancing the methods for transferring unknown unknowns to known unknowns during cyber threat intelligence analysis via structured analytic techniques, thereby narrowing down systematic biases and random noise.

Furthermore, the development of automated tools and techniques, leveraging the power of artificial intelligence (AI) and machine learning (ML), to support the application of TIBSA in practical scenarios would be highly beneficial in terms of efficiency, scalability and accuracy. AI and ML algorithms have shown great potential in analyzing large datasets, identifying patterns, and making predictions. By integrating AI and ML capabilities (like Bayesian inference) into the TIBSA methodology, it becomes possible to automate certain steps, such as data collection, threat analysis, and uncertainty modeling.

Lastly, TIBSA should be integrated with existing risk management frameworks and standards to provide a comprehensive approach to risk analysis and management. Future research should explore the compatibility and synergy between TIBSA and frameworks such as the NIST CSF v2 or ISO 27001/27005. This integration can enhance the interoperability and adoption of TIBSA within organizations.

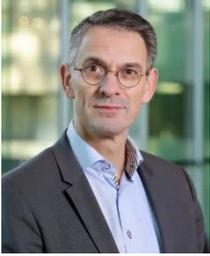
**Martijn Dekker** received the M.Sc. degree in mathematics at the University Utrecht in 1993 and the Ph.D. degree in mathematics at the University of Amsterdam in 1997. He has been an associate professor at the TIAS business school for Business and Society from 2012 to 2020. Since 2020 he is visiting professor Information Security at the University of Amsterdam / Amsterdam Business School. He is also the Chief Information Security Officer (CISO) at ABN AMRO.

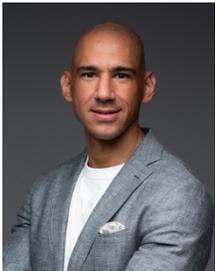
**Lampis Alevizos** received the M.Sc. degree in cybersecurity with the University of Central Lancashire (UCLan). He is currently pursuing the Ph.D. degree in Computer Science, researching ZTA, blockchain, and DLT convergence with cyber security. Additionally, he is a field Product Owner and a Senior Subject Matter Expert within ABN AMRO, solving daily cross-functional challenges. Lampis holds and actively maintains several industry certifications, such as CISSP, CCSP, CCSK, CISA, CISM and others.